\documentclass[floatfix,aps,10pt,prl,twocolumn,nopreprintnumbers,groupedaddress,superscriptaddress,tightenlines,longbibliography,letterpaper,altaffilletter,showpacs]{revtex4-2}

\usepackage[dvipsnames]{xcolor}
\usepackage{verbatim}
\usepackage{ulem}
\usepackage{bm,physics,mathrsfs}
\usepackage{enumerate}
\usepackage{epsf,amssymb,bbm,amsbsy,amsthm}
\usepackage{hyperref}
\usepackage{graphicx}
\usepackage{booktabs}
\usepackage{soul}
\usepackage{tikz,subfigure}
\usepackage{stackengine,scalerel}
\usepackage{BOONDOX-cal}
\usepackage[T1]{fontenc}
\usepackage{times}

\allowdisplaybreaks
\newcommand\dAlaux{%
  \Shortstack{\rule{12pt}{.6pt}\\
    \rule{.6pt}{10pt}\kern10pt\rule{1.4pt}{10pt}\\
    \rule{12pt}{1.4pt}}%
}
\newcommand\dAl{%
  \setstackgap{S}{0pt}%
  \setstackEOL{\\}%
  \scalerel*{\kern1pt\dAlaux\kern1pt}{\Delta}%
}

\definecolor{darkgreen}{RGB}{0,102,102}
\definecolor{purple}{RGB}{102,0,102}
\definecolor{darkblue}{RGB}{0,0,102}
\hypersetup{
    colorlinks=true,       
    linkcolor=purple,         
    citecolor=darkblue,        
    filecolor=magenta,     
    urlcolor=darkgreen         
}

\begin{document}
\title{Gravitational Instabilities of Uniform Black Strings in AdS}
\author{Aditya Dhumuntarao}
\email{dhumu002@umn.edu}
\affiliation{School of Physics and Astronomy, University of Minnesota Minneapolis, MN 55455, USA}
\author{Rafid Mahbub}
\email{mahbu004@umn.edu}
\affiliation{School of Physics and Astronomy, University of Minnesota Minneapolis, MN 55455, USA}

\begin{abstract}
	Locally AdS$_{d-1}\times\mathbb{R}$ uniform black strings (UBS) in the presence of a massless scalar field are believed to avoid the onset of the Gregory-Laflamme (GL) instability in $d\ge4$ as no tachyonic modes exist in the spectrum of the Laplace-Beltrami operator.  We present analytic and numerical evidence of GL modes in the Lichnerowicz spectrum indicating that AdS$_{d-1}$ UBSs are classically and thermodynamically unstable at the linear level in $d>4$. In $d=4$, we confirm that uniform BTZ$_3$ strings are indeed stable as previously suggested. We propose that linear instabilities of black strings are triggered if and only if a tachyonic mode exists in the Lichnerowicz spectrum. At the end state of the instability,  AdS$_{d-1}$ UBSs of finite length may tunnel to a SAdS$_d$ black hole or converge onto a novel non-uniform AdS$_d$ black string. We conjecture that weak cosmic censorship is violated if the non-uniform solution is an exact AdS$_d$ black funnel and compute entropy estimates in $d>4$ as evidence.
\end{abstract}
\maketitle
	\textit{Introduction.}\ Black objects with translationally invariant horizons suffer from classical instabilities. A famous example is the Gregory-Laflamme (GL) instability of Ricci-flat black strings in $d>4$, with horizon topologies $S^{d-3}\times S^1$, where tachyonic modes develop in the Lichnerowicz operator spectrum $(\Delta_\text{L})$ under generic long wavelength perturbations \cite{Gregory_1993,Gregory_1994,PhysRevD.66.064024,PhysRevD.73.104034}. These classical instabilities persist to four spacetime dimensions and higher \cite{Gregory_2000,PhysRevD.78.064001,Kang:2002hx,PhysRevD.64.064010,Kang:2004hm,PhysRevD.64.064010,Mann:2006yi,Brihaye:2007ju,PhysRevD.72.104019,Bernamonti:2007bu,Marolf:2019wkz,Bea:2020ees,Hubeny:2009rc,Hubeny:2009kz,PhysRevD.64.044005,Gubser:2000mm,Gubser:2000ec,Ross:2005vh} for black strings with asymptotics that are conformal to AdS$_{d-1}\times\mathbb{R}$. There is strong numerical evidence that GL modes lead to the horizon developing a finite time pinch-off -- constituting a violation of weak cosmic censorship \cite{lehner2011final}. For black strings with non-trivial matter configurations, tachyonic modes are also observed in the Laplace-Beltrami operator $(\Box)$. It is generally expected that the spectra of scalar operators on black string backgrounds are correlated. Thus finding tachyonic modes in the matter sector is a necessary and sufficient condition for classical instability and indeed, several numerical and analytic situations exist where such instabilities were only sourced from $\text{Spec}(\Box)$  \cite{PhysRevD.64.044005,Gubser:2000mm,Gubser:2000ec,Ross:2005vh}. 

	These linear instabilities are correlated to local thermodynamic instabilities. In its modern form, correlated stability states, for black strings with mass $M$ and conserved charges $Q_A$, classical instabilities are triggered precisely when a positive eigenvalue exists in the Hessian of the microcanonical entropy $S(M,Q_A)$ \cite{Gubser:2000mm,Gubser:2000ec}. For unique black strings with conserved charges, correlated stability has demonstrated that tachyonic modes develop in the spectrum of scalar operators precisely when the heat capacity becomes negative \cite{Gubser:2002yi,Bernamonti:2007bu,PhysRevD.67.024007,Kang:2002hx,PhysRevD.64.064010,Kang:2004hm,Buchel:2005nt,Miyamoto:2007mh,PhysRevD.78.064001,PhysRevD.78.126001}, and enjoys a general proof for asymptotically flat spacetimes \cite{Hollands:2012sf}.

	Recently, a new class of locally AdS$_{d-1}\times\mathbb{R}$ uniform black string (AdS$_{d-1}$ UBS) solutions in the presence of a massless scalar field has furnished two interesting puzzles \cite{Cisterna:2017qrb,Cisterna:2019scr}. First, no such tachyonic modes were identified in Spec$(\Box)$ leading to the conjecture that these solutions are linearly stable in four spacetime dimensions and higher \cite{Cisterna:2019scr}. This result is clearly in tension with earlier studies which have numerically and analytically confirmed a tachyonic mode in $\Delta_\text{L}$ for black strings which are locally conformal to AdS$_{d-1}\times\mathbb{R}$ \cite{PhysRevD.78.064001,Gregory_2000,PhysRevD.64.064010,Mann:2006yi,Brihaye:2007ju,PhysRevD.64.044005,Bernamonti:2007bu,Marolf:2019wkz,Bea:2020ees,Hubeny:2009rc,Hubeny:2009kz,PhysRevD.66.064024,PhysRevD.72.104019,PhysRevD.73.104034,Kang:2002hx,Kang:2004hm}. As the stability analysis performed in \cite{Cisterna:2017qrb,Cisterna:2019scr} primarily focused on Spec$(\Box)$, it is natural to ask \textit{does a GL mode persist in Spec$(\Delta_\text{L})$ in four spacetime dimensions and higher?}

	Secondly, the AdS$_{d-1}$ UBS is thermodynamically unstable when the radius of the string is sufficiently thin but remains classically stable in this regime. It is well known that non-unique black brane configurations with exotic scalar charge are counterexamples to correlated stability \cite{PhysRevD.72.104019}. However, the AdS$_{d-1}$ UBS is a unique configuration without a conserved charge associated with the massless scalar \cite{Cisterna:2017qrb,Cisterna:2019scr}, naturally raising the question \textit{are these solutions a new counterexample to the stronger form of correlated stability?}

	In this Letter, we present an analytic and numerical study of the classical and thermodynamic instability of locally AdS$_{d-1}\times\mathbb{R}$ uniform black strings coupled to a massless scalar. In concordance with the study initiated in \cite{Cisterna:2019scr}, we confirm that no GL modes exist in the Spec$(\Box)$. However, we find evidence of tachyonic modes in Spec$(\Delta_\text{L})$ in $d>4$ suggesting that the AdS$_{d-1}\times\mathbb{R}$ UBS is classically unstable. We also find a positive eigenvalue in Hess$(S)$ which, together with the GL modes in $\Delta_\text{L}$, is sufficient to rescue correlated stability in $d>4$. The reason the scalar operator spectra are uncorrelated may be understood by performing Kaluza-Klein (KK) compactifications along the flat direction. The resulting tower of masses for the scalar fluctuations satisfy the Breitenl\"ohner-Freedman bound on the locally AdS$_{d-1}$ background \cite{Breitenlohner:1982jf,Breitenlohner:1982bm}. Hence deformations from the matter sector do not grow sufficiently fast to trigger an instability, whereas the same is not true for propagating tensor modes in Spec$(\Delta_\text{L})$. To our knowledge, this is the first example where the spectra of scalar operators are uncorrelated for AdS black strings in the presence of non-trivial matter. This suggests tachyonic modes in Spec$(\Box)$ are sufficient, but not necessary as previously thought, to generate classical instabilities. We propose that \textit{linear black string instabilities are triggered if and only if there exists a tachyonic mode in Spec$(\Delta_{L})$.} 

	\textit{The Model.}\ We parameterize the AdS$_{d-1}$ UBS solution by
	\begin{align}
		&\mathcal{S} = \frac{1}{16\pi G_d}\int\qty[R - 2\Lambda_\text{bulk} - \frac{1}{2}(\partial\sigma)^2]\sqrt{-g}\dd^dx,\label{eqn:action1}\\
		&\dd s^2 = \frac{(d-2)}{(d-1)}\frac{L_d^2}{\ell_{d-1}^2}\qty[\dd z^2 -V(r)\dd t^2 +\frac{\dd r^2}{V(r)} + r^2\dd s^2_{S^{d-3}}],\nonumber\\
		&V(r)= 1 - {m}{r^{4-d}} + {r^2}{\ell^{-2}_{d-1}}, \hspace{.05in}\sigma(z) = \sqrt{2(d-2)} \qty({z}/{\ell_{d-1}})\nonumber.
	\end{align}
	The bulk and boundary AdS length scales are set via $\Lambda_\text{bulk} = -\frac{(d-1)(d-2)}{2L_d^2}$ and $\Lambda_\text{bdy} = -\frac{(d-2)(d-3)}{2\ell_{d-1}^2}$. We consider a finite string with $z\in \mathbb{I} = [-\pi L_z/2,\pi L_z/2]$ and restrict our attention to $d\ge4$. The UBS is a solution to the equations
	\begin{equation}\label{eqn:EOMs}
		R_{ab} = \frac{2\Lambda_\text{bulk}}{d-2}{g}_{ab} + \frac{1}{2}\partial_a\sigma\partial_b\sigma\hspace{.5in}\Box\sigma=0
	\end{equation}
	provided the massless scalar $\sigma$ satisfies Neumann boundary conditions. Clearly, the scalar modifies $\Lambda_\text{bulk}$ along the $z$ direction and additional flat directions $x_i$ may be appended at the expense of adding scalars $\sigma_i$ with identical asymptotics. Henceforth, the volume of the space transverse to the UBS horizon is unity, $(16\pi G_d)^{-1}\int_{S^{d-3}\times \mathbb{I}}\sqrt{\gamma}=1$ \footnote{This yields the normalization \begin{equation}G_d = \frac{\omega_{d-3}L_z}{8}\qty(\frac{d-2}{d-1})^{\frac{1}{2}(d-2)}\qty(\frac{L_d}{\ell_{d-1}})^{d-2}\label{eqn:norm}\end{equation} which is convenient to use when computing the thermodynamics}. 

	\textit{Correlated Stability.}\ On constant $z$ hypersurfaces, the geometry describes an SAdS$_{d-1}$ BH where the largest positive root of $V(r_+)=0$ designates the horizon. The mass and entropy may then be directly computed using standard formulas on the codimension one surface
	\begin{align}
		{M} &= (d-3)r^{d-4}_+\qty[E_{S^{d-3}}+\frac{r_+^2}{\ell_{d-1}^2}], \hspace{.2in}{S} = 4\pi{r_+^{d-3}}
	\end{align}
	where $E_{S^{d-3}}$ is the energy of the sphere which vanishes in $d=4$ and is unity otherwise. As noted in \cite{Cisterna:2019scr}, there is no conserved charge associated with the scalar. Hence, the thermodynamic potential is simply $S\equiv S(M)$ in the microcanonical ensemble and searching for positive eigenvalues in $\text{Hess}(S)$ is equivalent to identifying a regime where $S(M)$ is convex. However, for AdS black objects, there is no simple expression for the entropy \cite{Gubser:2000mm,Gubser:2000ec}. Rather, assuming strictly positive temperatures (which is reasonable for SAdS$_{d-1}$ black strings), the stability requirement may be rephrased in terms of finding negative eigenvalues in the Hessian of $M(S)$ which is simply the statement that the heat capacity becomes negative. We find in $d=4$, $M(S)=(S/4\pi\ell_{d-1})^2$ and in $d>4$,
	\begin{equation}
		M(S)=(d-3) \left(\frac{S}{4\pi}\right)^{\frac{d-4}{d-3}} \qty[1+ \frac{\left({S}/{{4\pi}}\right)^{\frac{2}{d-3}}}{\ell_{d-1}^2}].
	\end{equation}
	A routine computation of the heat capacity $C = T(\partial_{S}^{2}M)^{-1}$ where $T=V'(r_+)/4\pi$, shows that $M(S)$ is concave in $d>4$ for $r_+<r_+^c = \sqrt{(d-4)/(d-2)}\ell_{d-1}$. Hence, sufficiently thin UBSs with $S(r_+)<S(r_+^c) = 4\pi r_+^c$ are thermodynamically unstable in $d>4$. In $d=4$, the solution is a uniform BTZ$_3$ string with a mass gap $M\ge0$ separating the black string from the AdS$_3$ $(M=-1)$ background which has a strictly positive heat capacity \cite{Banados:1992wn,Frassino:2015oca}. Via correlated stability, this naively suggests tachyonic modes condense in Spec$(\Box,\Delta_\text{L})$ for horizons $r_+< r_+^c$ in $d > 4$ whereas the BTZ$_3$ UBS is stable regardless of the horizon size. Hence, the GL instability should have a lower critical dimension of $d=5$. Some evidence of this is observed for BTZ$_3$ black strings which are locally conformal to AdS$_{3}\times\mathbb{R}$ \cite{Kang:2002hx,PhysRevD.64.064010,Kang:2004hm}. In $4\ge d$, Einstein-(A)dS$_{d-1}$ gravity is topological. Thus metric perturbations will not contain propagating degrees of freedom and should reduce to a pure gauge transformation. We study these expectations analytically and numerically shortly.




	\textit{Linear Classical Stability.}\ We show that tachyonic modes in Spec$(\Box,\Delta_\text{L})$ are uncorrelated on the AdS$_{d-1}$ UBS background and the BTZ$_3$ UBS is stable. We consider perturbing the metric and scalar field around the background Eqn.~\eqref{eqn:action1} 
	\begin{align}\label{eqn:perturbations}
		g_{ab}(x^{a})&=\bar{g}_{ab}(x^{a})+\epsilon H_{ab}(x^{a})+\mathcal{O}(\epsilon^2)\\
		\sigma(x^{a})&=\bar{\sigma}(z)+\epsilon\Phi(x^{a})+\mathcal{O}(\epsilon^2)
	\end{align}
	where $(\bar{g}_{ab},\bar{\sigma})$ satisfies the equations of motion \eqref{eqn:EOMs}. To $\mathcal{O}(\epsilon)$, the perturbations on the AdS$_{d-1}$ UBS background satisfy
	\begin{align}
		&\Box\Phi \equiv -\frac{1}{V}\frac{\partial^{2}\Phi}{\partial t^2}+\frac{1}{r^{d-3}}\frac{\partial}{\partial r}\left( r^{d-3}V\frac{\partial\Phi}{\partial r} \right) + \frac{\partial^{2}\Phi}{\partial z^2}=0\label{eq:laplace}\\
		&\Delta_\text{L}H_{ab} = -\frac{4\Lambda}{d-2} H_{ab} + 2\, \partial_{(a}\bar{\sigma}\partial_{b)}\Phi\label{eqn:einsteinperturbations}
	\end{align}
	where the Lichnerowicz operator in the transverse gauge is
	\begin{align}\label{eq:lichnerowicz}
	\Delta_{\text{L}}H_{ab}\equiv \Box H_{ab}&+2\bar{R}_{acbd}H^{cd}-2\bar{R}^{c}_{(a}H^{{\color{white}c}}_{b)c}.
	\end{align} 
	Now, identifying tachyonic modes in Spec$(\Box,\Delta_\text{L})$ is tantamount to finding regular solutions $(H_{ab},\Phi)$ which have a growing mode instability \cite{Gregory_1993,Gregory_1994,Gregory_2000,PhysRevD.73.104034,PhysRevD.66.064024}. As the UBS solution is spherically symmetric and has a Killing isometry along the flat direction, we search for growing modes in the $s$-wave gauge with the expansion
	\begin{align}
		\Phi(x^{a})&=e^{im_nz+\Omega t}\phi(r)\label{eqn:ansatzphi},\\
		H_{ab}(x^a)&=e^{im_nz+\Omega t}\mqty( h_{tt} & h_{tr} & h_{tz}  & \bm{0} \\ h_{tr} & h_{rr} & h_{r z} & \bm{0} \\ h_{tz} & h_{r z} & h_{zz} & \bm{0} \\ \bm{0} & \bm{0}& \bm{0} & {\bm{h}_{ij}}   )\label{eqn:ansatzmetric}.
	\end{align}
	Here $m_n=({n}/{L_z})$, $\bm{h}_{ij}\dd\theta^i\dd\theta^j \equiv K(r) \dd s^2_{S^{d-3}}$, and $h_{\mu\nu}\equiv h_{\mu\nu}(r)$ along the directions transverse to $S^{d-3}$.  In the presence of matter, we are allowed to work in the transverse gauge $H^{ab}{}_{;b} =\frac{1}{2}H{}^{;a}$ where $H=\bar{g}^{ab}H_{ab}$. Lastly, in order for the initial value problem to be well posed, the domain of dependence will include $\mathscr{I}^+$ on the conformal boundary of the SAdS$_{d-1}$ BH when prescribing data \cite{Gregory_1993,Gregory_1994,Gregory_2000}.

	\begin{figure*}[!thbp]
		\centering
		\subfigure{\includegraphics[width=1\columnwidth]{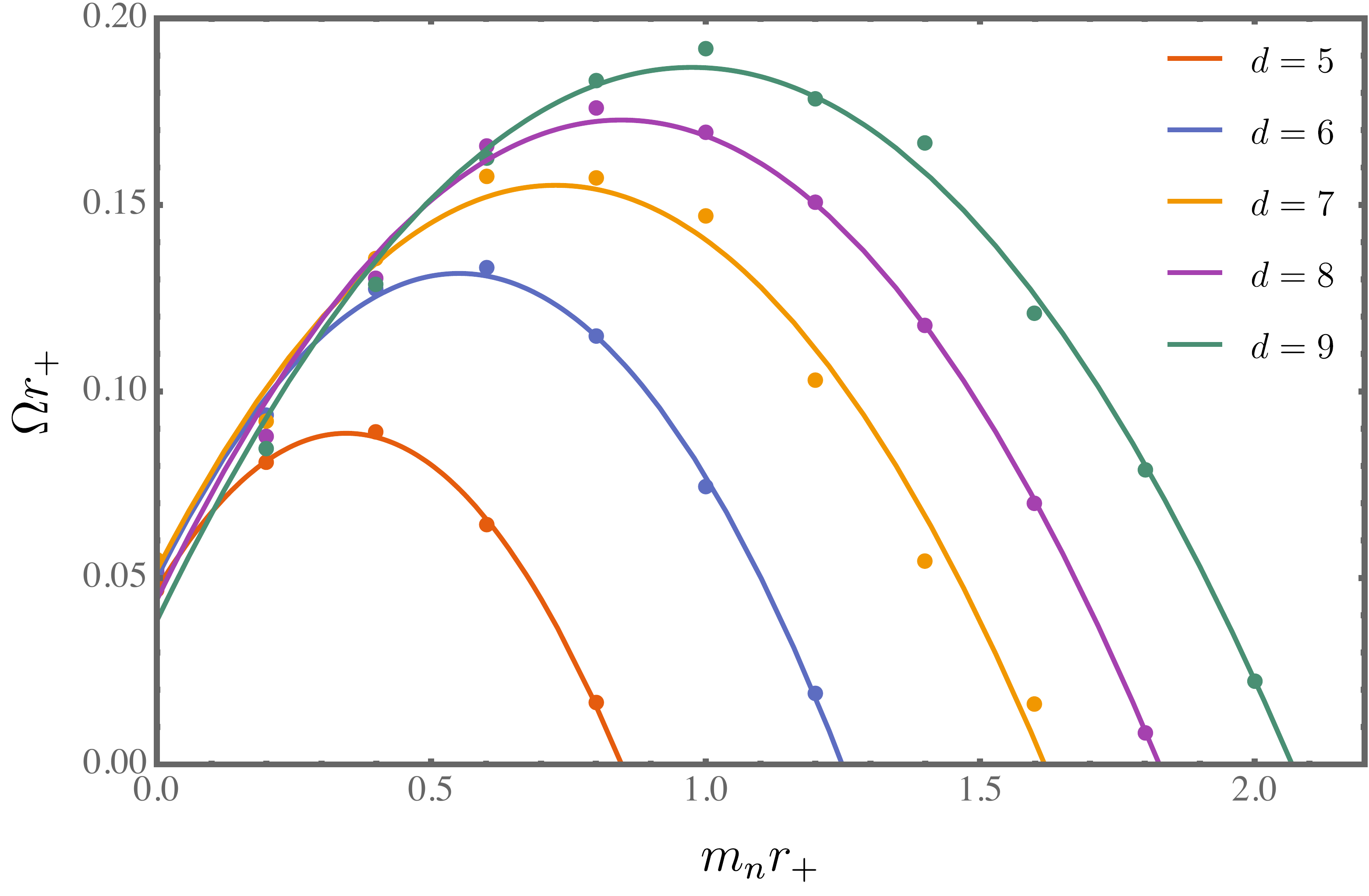}}\quad
  	\subfigure{\includegraphics[width=1\columnwidth]{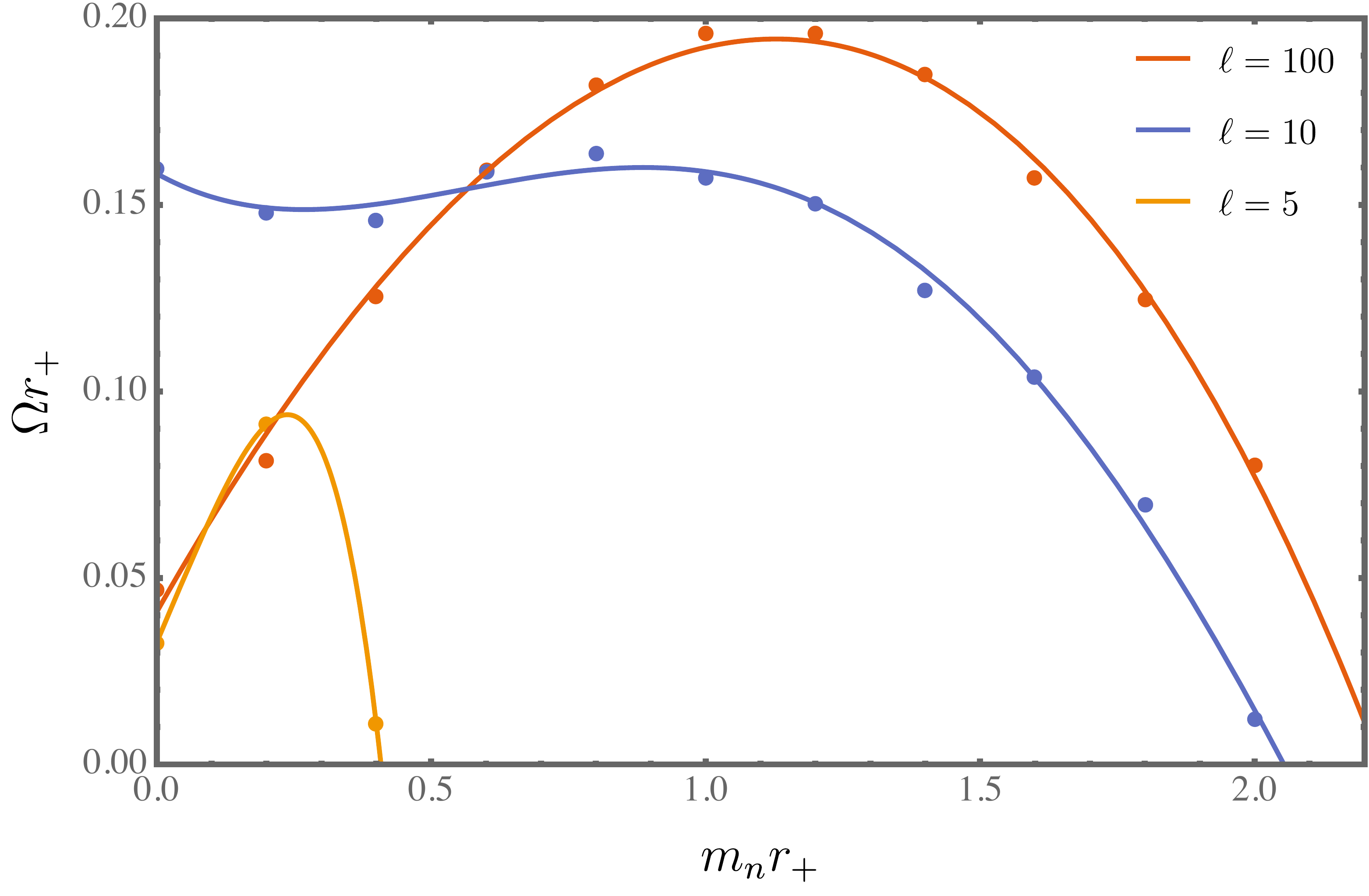}}
		\caption{Left: Plot of $(m_n,\Omega)$ with $r_+=2.0$ and $\ell = 50$ for which the ratio $\mathcal{R}$ changes sign in $9\le d\le5$. The points correspond to values calculated numerically indicating an instability. The lines are up-to-quadratic polynomial fits to illustrate the trend. Right: The instability in $d=10$ for various values of $\ell$. As $\mu_+\sim\mu_+^c$, the instability is terminated.}
		\label{fig:modeplot}
	\end{figure*}

	\textit{Matter Perturbations.} It is straightforward to show that no tachyonic modes may be found in Spec$(\Box)$ obeying regularity conditions near horizon $r\approx r_+$ and asymptotically $r\to\infty$  \cite{Cisterna:2019scr}. Evaluating Eqn.~\eqref{eq:laplace} on the ansatz Eqn.~\eqref{eqn:ansatzphi}, one finds
	\begin{align}\label{eq:radial_pert}
		V\phi''&+\left[ \frac{d-3}{r}V+V' \right]\phi' -\left[m_n^2 +\frac{\omega^2}{V} \right]\phi=0.
	\end{align}
	Regularity conditions enforce the behavior $\phi\sim (r-r_{+})^{\alpha}$ near horizon and $\phi\sim r^{-\beta}$ asymptotically with $\alpha,\beta\in\mathbb{R}^+$ \cite{PANI_2013}. As a result, the only way to achieve this behavior is if $\phi(r)$ possesses a maximum at some radial position $r_{\star}>r_+$. At the turning point, $\phi^\prime(r_\star)=0$ and hence we have
	\begin{equation}\label{eq:radial_pert2}
		V(r_{\star})\phi''(r_{\star})=\left[ m_n^2 +\frac{\omega^2}{V(r_{\star})} \right]\phi(r_{\star}).
	\end{equation}
	However, for this to be a maximum $V(r_{\star})\phi''(r_{\star})<0$ but no maximum may exist as $V(r_\star)>0$ for $r_\star>r_+$.

	We generalize the stability argument given in \cite{Cisterna:2019scr} by performing a KK compactification of the flat direction. One would then find a tower of masses $(\Box^{\text{SAdS}_{d-1}}-m_n^2)\Phi=0$ from which it is direct to see the Breitenl\"ohner-Freedman stability condition $m_n^2 = (n/L_z)^2 > -\frac{(d-2)^2}{4\ell_{d-1}^2}$ is satisfied \cite{Breitenlohner:1982jf,Breitenlohner:1982bm}. Hence, the AdS$_{d-1}$ UBS solution is indeed stable to scalar perturbations in $d\ge4$ independent of the coordinate gauge. As such, we eliminate this perturbation in our search for an instability \cite{Gregory_1994}. 

	\textit{Metric Perturbations.} The metric perturbations are quite involved. Nonetheless, it is possible to present the salient arguments, detailed in the appendix, which identify the growing mode instability. As the scalar perturbation decouples, the metric perturbations become tracefree $\bar{g}^{ab}H_{ab}=0$. Then gauge conditions, in conjunction with eliminating stable perturbations, can be used to reduce Eqn.~\eqref{eqn:einsteinperturbations} to a single ODE
	\begin{equation}\label{eqn:htrdfq}
		A_{m_n,\Omega}(r) h^{\prime\prime}_{tr}(r) + B_{m_n,\Omega}(r) h^{\prime}_{tr}(r) + C_{m_n,\Omega}(r)h_{tr}(r) = 0.
	\end{equation}
	
	The conditions on the radial boundary and the horizon are sensitive to the dimension and on $\Lambda_\text{bdy}$. The qualitative difference between $d=4$ and $d>4$ is best understood in the limit $(\Lambda_\text{bulk},\Lambda_\text{bdy})\to0$ where one is supposed to recover the instability of Schwarzschild$_{d-1}$ uniform black strings \footnote{We thank Matthew Headrick for bringing this to our attention}. In $d=4$, no Schwarzschild$_{3}$ UBS exist in the spectrum of Einstein gravity, where as the same is not true in $d>4$. Next, the dependence of the regularity condition on $\Lambda_\text{bdy}$ may be inferred in terms of the control parameter $\mu_+ = r_+/\ell_{d-1}$. For $\mu_+\gg1$, the radius of the AdS$_{d-1}$ UBS is large and requires sharply growing regular perturbations to destabilize the solution where as $\mu_+\ll1$ correspond to thin UBS solutions which are easier to perturb. We perform the parameter search in the latter regime in order to make contact with correlated stability. The conditions on the horizon and radial boundary may be found by solving Eqn.~\eqref{eqn:htrdfq} on the various asymptotics. We find
	\begin{align}
		r\to r_+ &: 
		\begin{cases}
			\mathcal{A}_\pm(m_n)(r-r_+)^{-1\pm\frac{r_+\Omega}{d-4}\sqrt{1-\varepsilon}},& d>4\\
			\mathcal{C}_\pm(m_n)(r-r_+)^{-1\pm\frac{r_+\Omega }{2\mu_+^2}},& d=4\\
		\end{cases}\\
		r\to\infty &: 
		\begin{cases}
			\mathcal{B}_\pm(m_n) e^{\pm m_n^2 r/\sqrt{m_n^2-\Omega^2}},&\hspace{.07in} d>4\\
			\mathcal{D}_{\pm}(m_n) e^{\pm\, \mathrm{i}\, m_n^2 r/\sqrt{\mu_+^2m_n^2+\Omega^2}},&\hspace{.07in} d=4\\
		\end{cases}
	\end{align}
	where $\varepsilon=\frac{(d-2)}{(d-4)}\mu_+^2$. In $d>4$, the near horizon condition suggests a real regular solution exists provided $\varepsilon<1$ and $\Omega< {(d-4)}/{r_+\sqrt{1-\varepsilon}}$. The condition on $\varepsilon$ is analytic evidence of correlated stability as the control parameter is bounded above precisely by the thermodynamic instability $\mu_+^2<\frac{(d-4)}{(d-2)}\equiv (r_+^c/\ell_{d-1})^2$. Near horizon, one recovers the familiar Schwarzschild$_{d-1}$ UBS behavior when $\mu_+=0$ \cite{Gregory_1993,Gregory_1994}. In the far horizon, the solution differs due to the conformal structure on the asymptotically AdS$_{d-1}$ surface. 

	In $d=4$, it is evident the perturbation is pure gauge -- the asymptotic solutions are oscillatory rather than smoothly damped. Near horizon, instabilities are exponentially damped out from the spectrum as regular solutions require $\Omega< \frac{2\mu_+^2}{r_+} \ll1$. We believe this behavior to be related to the lack of propagating gravitational degrees of freedom in lower dimensions. It is natural to understand this as analytic evidence suggesting the classical stability of BTZ$_3$ UBSs which agrees with the expectations provided from correlated stability.

	The goal is now to perform a parameter search on $(m_n,\Omega)$ which yields regular solutions, i.e., {tachyonic modes} in Spec$(\Delta_\text{L})$. We numerically integrate Eqn.~\eqref{eqn:htrdfq} (see the appendix) to compute the instability pairs $(m_{n},\Omega)$ using an adaptive Runge-Kutta-Fehlberg (RK45) routine in \textit{Mathematica}. A backwards integration is carried out between $r_{1}=200.0$ and $r_{2}=r_++10^{-5}$ with the seed solution set by the decaying branch $h_{tr}\sim e^{-m_{n}^2 r/\sqrt{m_{n}^2-\Omega^2}}$. The existence of an instability is detected with a sign change in the ratio $\mathcal{R}=\mathcal{A}_-/\mathcal{A}_+$ \cite{Gregory_1993,Gregory_1994} where
	\begin{equation}
		\mathcal{R}=\frac{\left( \frac{r_{+}\Omega}{d-4}\sqrt{1-\varepsilon}-1 \right)h_{tr}-(r-r_{+})h_{tr}'}{\left( \frac{r_{+}\Omega}{d-4}\sqrt{1-\varepsilon}-1 \right)h_{tr}+(r-r_{+})h_{tr}'}(r-r_{+})^{\frac{2r_{+}\Omega}{d-4}\sqrt{1-\varepsilon}}.
	\end{equation}
	
	\noindent We confirm the AdS$_{d-1}$ UBS solution is indeed unstable in $10\ge d\ge5$, and in $d=10$ we studied the threshold of this instability for various control parameters. We found no such instability in $d=4$ or beyond $r_+^c/\ell_{d-1}$; the results are summarized in Fig.~[\ref{fig:modeplot}].

	\textit{Comments on Weak Cosmic Censorship Violation}.\ Naturally, one may ask which geometry arises at the end state of this instability, an Schwarzschild-AdS$_{d}$ black hole (SAdS$_{d}$ BH) or a novel non-uniform AdS$_{d}$ black string (AdS$_{d}$ nUBS) configuration, and if a violation of weak cosmic censorship occurs during the transition, especially in $d=4$. As it turns out, the existence of a GL mode is insufficient to provide a violation of weak cosmic censorship \footnote{We would like to thank Jorge Santos for valuable communications on this matter.} (see \cite{Marolf:2019wkz,Bea:2020ees} where GL points exist but a tunneling event is not expected to occur). One is further required to compute the entropy of the nUBS solution and determine its dominance in the phase diagram. It is then reasonable to conjecture that weak cosmic censorship is violated, if $(i)$ a GL mode exists, and $(ii)$ the non-uniform solution has a subdominant entropy relative to the uniform solution $(S_\text{nUBS}<S_\text{UBS})$.

\begin{figure}[!thbp]
		\centering
		\includegraphics[width=\columnwidth]{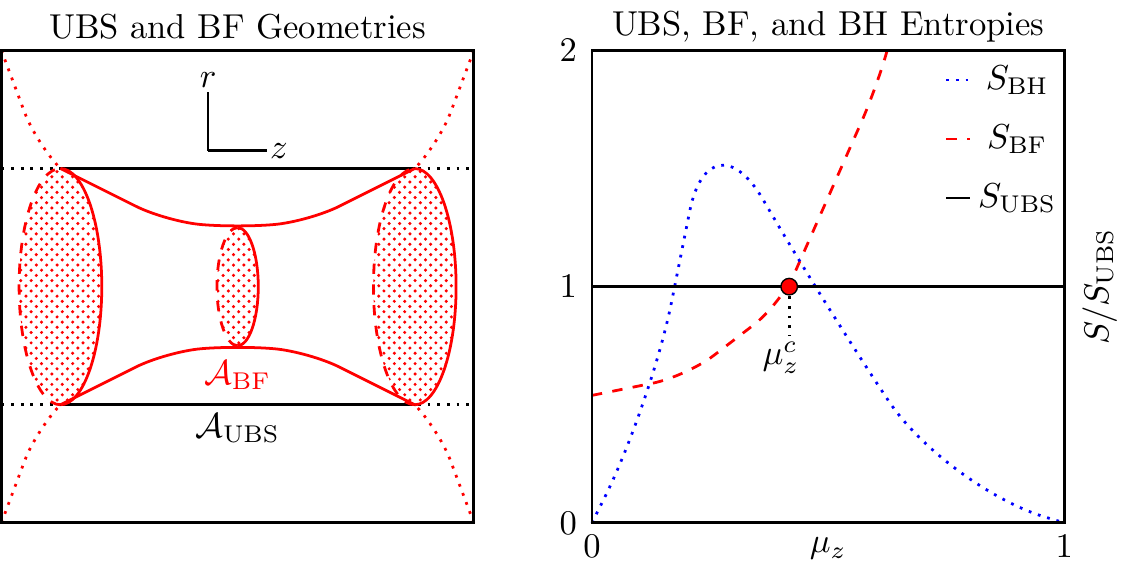}
		\caption{Left: Diagram of coincident UBS and BF geometries within the strip $z\in[-\pi L_z/2 ,\pi L_z/2]$ at equal mass. Right: The UBS is entropically dominant over the BF solution for $\mu_z < \mu_z^c$. The BH entropy is observed to follow the above trend.}
		\label{fig:entropyplot}
	\end{figure}

	In the absence of a dynamical scalar, a suggestive set of non-uniform saddles can be constructed by foliating AdS$_{d}$ in terms of AdS$_{d-1}$ slices and then replacing each slice with SAdS$_{d-1}$ black holes (BH). The resulting solutions to Eqn.~\eqref{eqn:action1} are exact AdS$_d$ black funnels (BF)  \cite{Hubeny:2009rc,Hubeny:2009kz}
	\begin{align}
		&\dd s^2_\text{BF} = \frac{(L_{d}/\ell_{d-1})^2}{\cos(z/\ell_{d-1})^2}\qty[\dd z^2 -V(\hat{r})\dd \hat{t}^2 + \frac{\dd \hat{r}^2}{V(\hat{r})}+\hat{r}^2\dd s_{S^{d-3}}^2],\nonumber\\
		& V(\hat{r}) = 1 - \hat{m}\hat{r}^{4-d} - \hat{r}^2\ell_{d-1}^2 ,\hspace{.1in}
		\sigma(z)=\sigma_0
	\end{align}
	which are locally conformal to AdS$_{d-1}\times\mathbb{R}$. The phase diagram now contains ({at least} \cite{Marolf:2019wkz}) three black objects with distinct horizon topologies $\mathcal{H}$ and scalar minima; SAdS$_{d}$ BH with $(\sigma=\sigma_0,\mathcal{H}=S^{d-2})$, AdS$_{d}$ BF with $(\sigma=\sigma_0,\mathcal{H}=\Omega^2_z({S}^{d-3}\times\mathbb{R}))$, and AdS$_{d-1}$ UBS with $(\sigma\propto z, \mathcal{H}=S^{d-3}\times\mathbb{R})$. On thermodynamic grounds, transitions between these geometries is expected at a critical point $M_\star$ \cite{Marolf:2019wkz}. Clearly, a violation of weak cosmic censorship is expected in any first order transition between black string and black hole phases, whereas fragmentation does not occur during transitions between black string phases which share conformally equivalent horizon topologies. If the BF phase is entropically dominant, then an energy barrier will separate the UBS and BH phases. Hence, the statement of weak cosmic censorship violation amounts to finding regimes where $S_\text{BF}<S_\text{UBS}$ so that transitions between black strings and black holes occur. Such a regime
	will depend on the control parameter $\mu_z = L_z/\ell_{d-1}$. One should expect that area difference of the UBS and BF strips, with length $z\in[-\pi L_z/2,\pi L_z/2]$, should be negative for sufficiently small $\mu_z$ (see Fig.~\ref{fig:entropyplot}). 
	We confirm this via entropy estimates in $d\ge 5$. The thermodynamics read
	\begin{align}
		M_\text{BF}&= {(d-3)}\hat{r}_+^{d-4}\qty[E_{S^{d-3}}+\frac{\hat{r}_+^2}{\ell_{d-1}^2}]I_d(\mu_z),\\
		S_\text{BF}&= 4\pi \hat{r}_+^{d-3}I_d(\mu_z),
	\end{align}
	where $I_d(\mu_z) = (2\pi)^{-1}(\frac{d-1}{d-2})^{(d-2)/2}\int_{-\pi/2}^{\pi/2}\cos(\mu_z x)^{2-d}\dd x$ which is convergent for $\mu_z<1$ (see Eqn.~\eqref{eqn:norm}). Along the line $M_\text{BF}=M_\text{UBS}$, the entropy difference may be estimated in the stable regime $(\mu_+,\hat{\mu}_+)\gg E_{S^{d-3}}$ via
	\begin{equation}
		\Delta S(\mu_z) = S_\text{UBS}-S_\text{BF} \approx S_\text{UBS}\qty(1- I_d(\mu_z)^{\frac{1}{d-2}}).
	\end{equation}
	The UBS and BF phases exchange dominance when $I_d(\mu_z^c)=1$. Under these conditions, it is reasonable to conjecture that \textit{weak cosmic censorship is violated provided the non-uniform saddle corresponds to an exact AdS$_{d}$ black funnel geometry.}

	\textit{Conclusions}.\  We find the spectra of scalar operators on Locally AdS$_{d-1}\times\mathbb{R}$ uniform black strings supported by a massless scalar \cite{Cisterna:2019scr,Cisterna:2017qrb} are indeed uncorrelated in $d\ge5$, contrary to prior expectations \cite{PhysRevD.64.044005,Gubser:2000mm,Gubser:2000ec,Ross:2005vh}, and this solution falls into the class of black strings covered by correlated stability in $d\ge4$. Tachyonic modes in the Laplace-Beltrami spectrum appear to be only sufficient to generate classical instabilities. We propose that black string instabilities are triggered iff there exists a tachyonic mode in Spec$(\Delta_\text{L})$. Lastly, we have found a regime of intermediate AdS$_{d-1}$ UBS solutions which are entropically favorable over exact AdS$_d$ BF and compete directly with Schwarzschild-AdS$_d$ BHs. Although a complete numerical study of the Einstein-Dilaton-AdS$_d$ model is needed to elucidate if weak cosmic censorship is violated, the thermodynamic argument presents some preliminary evidence that the instability can tunnel a UBS to a BH. In AdS$_{d}$ with a constant scalar, it is well known that additional black objects, such as black droplets, exist in the thermodynamic phase space \cite{Marolf:2019wkz}. These black objects do not change the qualitative argument given here as the horizons in such cases are partitioned, and hence it is reasonable to conjecture a violation of weak cosmic censorship in $d\ge5$. It is of interest to understand which configuration the AdS$_{d-1}$ UBS will ultimately converge onto.

{\it Acknowledgments.} 
The authors acknowledge the use of the \textbf{diffgeo} Mathematica package for their symbolic tensor calculations \cite{diffgeo}. We thank Jorge Santos, Robie Hennigar, Robert Mann, and Matthew Headrick for discussions and hospitality. We are supported by the University of Minnesota Doctoral Dissertation Fellowship. A.\ D.\ is supported by the National Science Foundation Graduate Research Fellowship under Grant No. 00039202.

{\it Note Added.} After the completion of the present analysis, we were made aware of similar work carried out independently in $d=5$ \cite{henriquezbaez2021stability}. The analysis of the AdS$_4$ UBS solution there involved a power series expansion of the metric perturbation $h_{tr}$ in order to search for an instability. We would like to thank the authors of \cite{Cisterna:2019scr} for sharing their preprint along with their helpful insights on the subject.

\appendix
\section{App.A: Metric Perturbations}\label{app:Metpert}
After the scalar perturbations decouple, the metric perturbation become trace free $\bar{g}^{ab}H_{ab}=0$ which we may use to set the function $K(r)$. Performing the KK compactification allows us to decompose the metric perturbations into a scalar part $h_{zz}$, a vector part $h_{\mu z}$, and a tensor part $h_{\mu\nu}$. It is straightforward to show that no unstable modes exist within the scalar and vector sectors \cite{Gregory_1994,Gregory_2000} which meet the criteria of being well behaved at the radial boundary and the future event horizon  (see \cite{PhysRevD.66.064024} for an indepth discussion). The metric perturbations in Eqn.~\eqref{eqn:einsteinperturbations} then consists of linear, coupled, second order ODEs for $(h_{tt},h_{rr},h_{tr})$. By taking the combination $h_\pm = \frac{h_{tt}}{V} \pm V{h_{rr}}$, we find an algebraic relation for $h_+$ and two first order linear ODEs for the perturbations $(h_-,h_{tr})$ 
\begin{widetext}
		\begin{align}
			h_+ \frac{V}{2}& \qty[m_n^2+\frac{(d-4) (d-3) (1-V)}{2 r^2}-\frac{(d+1) \Lambda_\text{bdy} }{d-3}]=\\
			&{\color{white}+}h_- \qty[\Omega ^2+\frac{m_n ^2 V}{2}+\frac{(d-4)}{4 r^2}\left(V^2+(d-5)V-(d-4)\right)+\frac{\Lambda_\text{bdy}}{(d-3)}\qty((d-4)-\frac{\Lambda_\text{bdy} r^2}{(d-3)}-\frac{(d+3) V}{2})]\nonumber\\
			&+h_{tr}\frac{ V }{r}\qty[\frac{2 (d-4) \Lambda_\text{bdy} }{(d-3) \Omega }-\frac{4 \Lambda_\text{bdy} ^2 r^2}{(d-3)^2 \Omega }+\frac{\Lambda_\text{bdy}  m_n ^2 r^2}{(d-3) \Omega }-\frac{2 (d-2) \Lambda_\text{bdy}  V}{(d-3) \Omega }-\frac{m_n ^2 (-(d-2) V+d-4)}{2 \Omega }-(d-3) \Omega ]\nonumber\\
		h_-'(r)&=\frac{h_{tr}}{\Omega } \left(m_n ^2-\frac{4 \Lambda_\text{bdy} }{d-3}\right)+\frac{h_-}{2 r V} \left((d-4)+(5-2 d) V-\frac{2 \Lambda_\text{bdy}  r^2}{d-3}\right)+\frac{(d-3) h_+}{2 r}\\
		h_{tr}'&=\left(h_-+h_+\right) \frac{\Omega }{2 V}-\frac{h_{tr}}{r V} \left((d-4)+V-\frac{2 \Lambda_\text{bdy}  r^2}{d-3}\right)
	\end{align}
\end{widetext}
	As a consistency check, the limit where $(\Lambda_\text{bulk},\Lambda_\text{bdy})\to0$ reduces the system $(h_+,h_-,h_{tr})$ to \cite{Gregory_1994} and we are able to reproduce the classical instability of Schwarzschild black strings. Via the scaling $r\to\ell_{d-1}u$, $\Omega \to \ell_{d-1}^{-1}\omega$, $m_n\to\ell_{d-1}^{-1} k_n$, and after some algebraic manipulations, we find a single second order ODE for $h_{tr}$ of the form
	\begin{widetext}
		\begin{align}\label{eq:htr_fullform}
			&\left[-2 V(u) \left(u^2 \left(d^2+2 k_n^2-4\right)+(d-4)^2\right)+\left((d-2) u^2+d-4\right)^2+(d-4)^2 V(u)^2-4 u^2 \omega ^2\right]H''(u) \\
			+&\bigg[V(u) \left(u^2 \left(4 (d-5) k_n^2+(d-2) (d (3 d-14)-4)\right)+3 (d-4)^2 V(u)+3 (d-6) (d-4)^2\right) +4 (2 d-9) u^2 \omega ^2\nonumber\\
			&-\left((d-2) u^2+d-4\right) \left(u^2 \left((d-2) (6 d+1)+8 k_n^2\right)+3 (d-4) (2 d-9)\right)\nonumber\\
			&+\frac{3 \left((d-2) u^2+d-4\right) \left(\left((d-2) u^2+d-4\right)^2-4 u^2 \omega ^2\right)}{V(u)}\bigg]\frac{H'(u)}{u}\nonumber\\
			+&\bigg[V(u) \bigg\{V(u) \bigg(V(u) \left(u^2 \left(\left(d^2-12\right) k_n^2+(d-2) (d ((d-5) d+14)-28)\right)+(d-4)^2 V(u)+((d-5) d+2) (d-4)^2\right)\nonumber\\
			&+u^2 \left(((44-5 d) d-92) \omega ^2-2 (d-4) (d-2) \left(d^2+k_n^2-11\right)\right)\nonumber\\
			&+u^4 \left(-2 (d-8) (d-2) k_n^2-(d-2)^2 (d (d+2)-13)+4 k_n^4\right)-(d-4)^2 ((d-2) d-9)\bigg)\nonumber\\
			&+2 u^2 \omega ^2 \left(u^2 \left((d-2) (5 d-14)+4 k_n^2\right)+5 (d-4)^2\right)+\left((d-2) u^2+d-4\right)^2 \left(u^2 k_n^2-(d-7) \left((d-2) u^2+d-4\right)\right)\bigg\}\nonumber\\
			&-5 u^2 \omega ^2 \left((d-2) u^2+d-4\right)^2+\left((d-2) u^2+d-4\right)^4+4 u^4 \omega ^4\bigg]\frac{H(u)}{u^2 V(u)^2}\nonumber=0
		\end{align}
	\end{widetext}
	where $V(u) = 1-(1+u_+^2)(u_+/u)^{d-4} + u^2$ and $u_+\ll1$ is used in the numerical simulation.

\bibliography{DMReferences}
\end{document}